\documentstyle[aps,twocolumn]{revtex}

\begin{document}
\twocolumn[\hsize\textwidth\columnwidth\hsize\csname@twocolumnfalse\endcsname

\title
{Greenberger-Horne-Zeilinger nonlocality for continuous variable systems}

\author{Zeng-Bing Chen$^{1}$ and Yong-De Zhang$^{2,1}$}
\address
{$^1$Department of Modern Physics, University of Science and Technology of China,
Hefei, Anhui 230027, P.R. China}
\address
{$^2$CCAST (World Laboratory), P.O. Box 8730, Beijing 100080, P.R. China}
\date{\today}
\maketitle 

\begin{abstract}

\
As a development of our previous work, this paper is concerned with
the Greenberger-Horne-Zeilinger (GHZ) nonlocality for continuous variable cases.
The discussion is based on the introduction of a pseudospin operator, which has the same algebra
as the Pauli operator, for each of the $N$ modes of a light field. Then the Bell-CHSH
(Clauser, Horne, Shimony and Holt) inequality is presented for the $N$ modes, each
of which has a continuous degree of freedom. Following Mermin's argument, it is
demonstrated that for $N$-mode parity-entangled GHZ states (in an infinite-dimensional
Hilbert space) of the light field, the contradictions between quantum mechanics and
local realism grow exponentially with $N$, similarly to the usual $N$-spin cases.

\

PACS number(s): 03.65.Ud, 03.65.Ta

\

\

\end{abstract}]

Under the presumption of local realism, Einstein, Podolsky and Rosen (EPR)
argued that it seems possible to assign values with certainty to canonically
conjugate variables (e.g., position and momentum) of one particle by
performing experiments on another particle of an entangled biparticle system 
\cite{EPR}. EPR's observation is definitely in conflict with quantum
mechanics. The entangled states of the biparticle system are now known as
the EPR states of continuous variables. In 1965 Bell, using Bohm's EPR
states of discrete variables (e.g., spin) \cite{Bohm}, derived the famous
Bell's inequalities \cite{Bell,CHSH,Bell-book}, enabling quantitative tests
of quantum mechanics versus local realism. So far, many experiments based on
Bohm's EPR states completely confirmed quantum mechanics \cite{Aspect}.
Bell's theorem without inequalities has also been demonstrated for
nonmaximally entangled biparticle states \cite{Hardy} and multiparticle
Greenberger-Horne-Zeilinger (GHZ) states \cite{GHZ-89,GHZ-90}. In this case,
the contradictions between quantum mechanics and local realistic theories
are of a non-statistical nature; the nonlocality can be manifest in a single
measurement. Using three-photon polarization-entangled states, Pan {\it et al%
}. carried out an experimental test of GHZ theorem \cite{Pan-GHZ},
confirming again the correctness of quantum mechanics. Besides its
theoretical interest, research work on quantum nonlocality also gains great
impetus from the rapidly developing field of quantum information theory \cite
{QIT}.

Quantum nonlocality for continuous variables has also attracted much
attention recently \cite{Bell-book,Bell86,Banaszek,Chen}. Several proposals
have been put forward to prepare the original EPR states \cite
{Reid,Walls-book,Parkins}; EPR paradox in EPR's original sense has been
realized experimentally \cite{Ou,Ou-APB}. The experiment exploited the
nondegenerate optical parametric amplifier (NOPA), which generates the
two-mode squeezed vacuum states (``regularized'' EPR states) 
\begin{equation}
\left| {\rm NOPA}\right\rangle =\frac 1{\cosh r}\sum_{n=0}^\infty (\tanh
r)^n|n,n\rangle ,  \label{nopa}
\end{equation}
with $r>0$ denoting the squeezing parameter and $\left| nn\right\rangle
\equiv \left| n\right\rangle _1\otimes \left| n\right\rangle _2$ the
twin-photon states. In the limit of infinite squeezing, the states $\left| 
{\rm NOPA}\right\rangle $ reduce to a normalized version of the original EPR
states \cite{Ou,Ou-APB}. Since the Wigner function \cite{Wigner} of $\left| 
{\rm NOPA}\right\rangle $ is positive everywhere, it has been argued that
the original EPR states may allow a hidden variable description and thus
will not exhibit nonlocality \cite{Bell-book,Bell86,Ou,Ou-APB}. This point
of view was dramatically changed by Banaszek and W\'odkiewicz (BW) \cite
{Banaszek}. Within a phase space formulation, BW used the ``Bell operator'' 
\cite{Bell-op} based on the joint parity measurements and demonstrated the
violations of the Bell inequality due to Clauser, Horne, Shimony and Holt
(CHSH) \cite{CHSH}. GHZ nonlocality for continuous variables has also been
analyzed \cite{GHZ-cv}, relying on the BW formalism. On the experimental
aspect, a recent experiment \cite{exp} observed, for the first time, the
violations of the Bell-CHSH inequality by the regularized EPR states
produced in a pulsed NOPA within the homodyne detection scheme.

In a recent paper \cite{Chen}, we have generalized Bell's inequalities to
the continuous variable cases for biparticle systems. Our formalism is based
on a new Bell operator, which is a direct analogy of its discrete variable
counterpart \cite{Bell-op}. In such a physically appealing manner, Bell's
inequalities for both discrete variable (in a two-dimensional Hilbert space)
and continuous variable (in an infinite-dimensional Hilbert space) cases
take a mathematically similar form. In this paper, we extend the work to
entangled multiparty systems. In particular, we will present GHZ theorem for
the parity-entangled GHZ states of a three-mode light field. Moreover, the
Bell-CHSH inequality for multiparty systems of discrete variables \cite
{Bell-N} will be generalized to continuous variable cases.

Let us start with considering a light field with three independent modes $1$%
, $2$ and $3$. For each mode of the light field, we can introduce the
``parity spin'' operator (i.e., pseudospin operator acting upon the parity
space of photons) \cite{Chen} 
\begin{eqnarray}
{\bf \hat s}_j &=&(s_{jx},s_{jy},s_{jz}),\;\;\;(j=1,2,3)  \nonumber \\
s_{jx} &=&s_{j+}+s_{j-},\;\;s_y=-i(s_{j+}-s_{j-}).  \label{sv}
\end{eqnarray}
Here 
\begin{eqnarray}
s_{jz} &=&\sum_{n=0}^\infty \left[ \left| 2n\right\rangle _{jj}\left\langle
2n\right| -\left| 2n+1\right\rangle _{jj}\left\langle 2n+1\right| \right] , 
\nonumber  \label{oe} \\
s_{j+} &=&\sum_{n=0}^\infty \left| 2n\right\rangle _{jj}\left\langle
2n+1\right| =(s_{j-})^{\dagger }  \label{s}
\end{eqnarray}
with $\left| n\right\rangle _j$ denoting the usual Fock states of the $j$%
-mode light field. After a simple algebra, it can be shown that 
\begin{eqnarray}
&&\left. \lbrack s_{iz},s_{j\pm }]=\pm 2\delta _{ij}s_{i\pm
},\;\;\;\;[s_{i+},s_{j-}]=\delta _{ij}s_{iz}.\right.  \nonumber  \label{comm}
\\
&&\left. s_{i\mu }s_{j\nu }+s_{i\mu }s_{j\nu }=2\delta _{ij}\delta _{\mu \nu
}.\;\;\;(\mu ,\nu =x,y,z)\right.  \label{comm}
\end{eqnarray}
The commutation relations in Eq. (\ref{comm}) are identical to their
counterparts of the spin-$1/2$ systems. Thus each parity spin operator ${\bf 
\hat s}_j$ acts as the counterpart of the Pauli operator ${\bf \sigma }$. In
particular the following relation is valid 
\begin{eqnarray}
U_j(\zeta _j,{\bf \hat n}_j) &=&\exp \left( -i\frac{\zeta _j}2{\bf \hat n}%
_j\cdot {\bf \hat s}_j\right)  \nonumber \\
&=&\cos \frac{\zeta _j}2-i{\bf \hat n}_j\cdot {\bf \hat s}_j\sin \frac{\zeta
_j}2  \label{rotate}
\end{eqnarray}
for ${\bf \hat s}_j$ as for ${\bf \sigma }$. Here ${\bf \hat n}_j$ is an
arbitrary unit vector. $U_j(\zeta _j,{\bf \hat n}_j)$ are, in fact, the
rotation operators in the parity space of photons.

Now one can introduce states of definite parity for each mode 
\begin{equation}
\left| +\right\rangle _j\equiv \sum_{n=0}^\infty {\cal A}_n^{(j)}\left|
2n\right\rangle _j,\;\;\left| -\right\rangle _j\equiv \sum_{n=0}^\infty 
{\cal A}_n^{(j)}\left| 2n+1\right\rangle _j,  \label{par-state}
\end{equation}
where $\sum_{n=0}^\infty \left| {\cal A}_n^{(j)}\right| ^2=1$. Some
important properties of these parity states are 
\begin{eqnarray}
s_{jz}\left| \pm \right\rangle _j &=&\pm \left| \pm \right\rangle _j, 
\nonumber \\
s_{j\pm }\left| \pm \right\rangle _j &=&0,\;\;\;s_{j\pm }\left| \mp
\right\rangle _j=\left| \pm \right\rangle _j.  \label{prop}
\end{eqnarray}
Hence $\left| \pm \right\rangle _j$ are the eigenstates of the parity
operators $s_{jz}$, with the corresponding eigenvalues $\pm 1$; the
operators $s_{j\pm }$ ``flip'' the parity of $\left| \mp \right\rangle _j$
and annihilate $\left| \pm \right\rangle _j$. In this aspect, $\left| \pm
\right\rangle _j$ are similar to the spin-up and spin-down states of the
spin-$1/2$ systems. Using the rotation operators as in Eq. (\ref{rotate}),
one can construct eigenstates of ${\bf \hat n}_j\cdot {\bf \hat s}_j$ for
each mode. Now the parity-entangled GHZ state of the light field can be
directly constructed as 
\begin{equation}
\left| {\rm GHZ}\right\rangle _3=\frac 1{\sqrt{2}}\left( \left|
+\right\rangle _1\left| +\right\rangle _2\left| +\right\rangle _3-\left|
-\right\rangle _1\left| -\right\rangle _2\left| -\right\rangle _3\right) .
\label{ghz}
\end{equation}
Note that this state is defined in an infinite-dimensional Hilbert space.

Following GHZ's argument \cite{GHZ-89,GHZ-90,QIT}, it is directly verifiable
that the GHZ state (\ref{ghz}) satisfies the eigenvalue equations as 
\begin{eqnarray}
s_{1x}s_{2x}s_{3x}\left| {\rm GHZ}\right\rangle _3 &=&-\left| {\rm GHZ}%
\right\rangle _3,  \nonumber \\
s_{1x}s_{2y}s_{3y}\left| {\rm GHZ}\right\rangle _3 &=&\left| {\rm GHZ}%
\right\rangle _3,  \nonumber \\
s_{1y}s_{2x}s_{3y}\left| {\rm GHZ}\right\rangle _3 &=&\left| {\rm GHZ}%
\right\rangle _3,  \nonumber \\
s_{1y}s_{2y}s_{3x}\left| {\rm GHZ}\right\rangle _3 &=&\left| {\rm GHZ}%
\right\rangle _3.  \label{ghz4}
\end{eqnarray}
Thus quantum mechanically, one can determine any operator ($s_{1x}$, $s_{1y}$%
, $s_{2x}$,$\ldots $) by performing appropriate measurements on the other
two modes of the light field. To establish a local realistic interpretation
of the quantum-mechanical results (\ref{ghz4}), one assumes that the
individual value of any operator ($s_{1x}$, $s_{1y}$, $s_{2x}$,$\ldots $) is
predetermined, e.g., by local hidden variables, regardless of any set of
parity spin measurements on the three modes of the light field. These
predetermined values are denoted by $p_{1x}$, $p_{1y}$, $p_{2x}$,$\ldots $,
with $p_{jx},p_{jy}=\pm 1$. To be consistent with Eq. (\ref{ghz4}), local
realistic theories impose the following relations 
\begin{eqnarray}
p_{1x}p_{2x}p_{3x} &=&-1,\;\;p_{1x}p_{2y}p_{3y}=+1,  \nonumber \\
p_{1y}p_{2x}p_{3y} &=&+1,\;\;p_{1y}p_{2y}p_{3x}=+1.  \label{hv}
\end{eqnarray}
But in fact, these relations are not mutually consistent. For example, the
latter three relations in Eq. (\ref{hv}) give $%
p_{1x}p_{2x}p_{3x}p_{1y}^2p_{2y}^2p_{2y}^2=p_{1x}p_{2x}p_{3x}=+1$, which
conflicts with the first relation in Eq. (\ref{hv}). Here we have used the
fact that $p_{jy}^2=1$. Thus quantum-mechanical predictions are incompatible
with local realistic theories for the parity-entangled GHZ states (\ref{ghz}%
). Also, the contradictions between quantum mechanics and local realism
occur for perfect correlations (i.e., for non-statistical predictions) of
the three-mode light field. It is worth pointing out that the GHZ paradox
associated with position-momentum variables has also been considered in
terms of the Weyl algebra in Ref. \cite{Clifton}.

In a real experiment, the perfect correlation condition in the GHZ argument
is very difficult to implement practically. To face this difficulty, one has
to rely on the multiparty generalization of Bell's inequalities, which has
been derived originally for discrete variables \cite{Bell-N}. Here we
generalize the multiparty Bell-CHSH inequality to the present continuous
variable case.

First, let us define the Bell operator in the case of a two-mode light field 
\cite{Chen}: 
\begin{eqnarray}
{\cal B}_2 &=&({\bf a}_1\cdot {\bf \hat s}_1)\otimes [({\bf a}_2\cdot {\bf 
\hat s}_2)+({\bf a}_2^{\prime }\cdot {\bf \hat s}_2)]  \nonumber \\
&&\ \ +({\bf a}_1^{\prime }\cdot {\bf \hat s}_1)\otimes [({\bf a}_2\cdot 
{\bf \hat s}_2)-({\bf a}_2^{\prime }\cdot {\bf \hat s}_2)],  \label{bellop}
\end{eqnarray}
where ${\bf a}_{1,2}$ and ${\bf a}_{1,2}^{\prime }$ are four unit vectors.
The commutation relations in Eq. (\ref{comm}) lead to, e.g., 
\begin{equation}
({\bf a}_1\cdot {\bf \hat s}_1)^2=I_1,  \label{unit}
\end{equation}
where $I_1$ is the unit operator for the mode-$1$ of the light field.
Equation (\ref{unit}) implies that the eigenvalues of the Hermitian operator 
${\bf a}_1\cdot {\bf \hat s}_1$ are $\pm 1$. Now the $N$-mode Bell operator
can be recursively defined by 
\begin{eqnarray}
{\cal B}_N &=&{\cal B}_{N-1}\otimes \frac 12[({\bf a}_N\cdot {\bf \hat s}%
_N)+({\bf a}_N^{\prime }\cdot {\bf \hat s}_N)]  \nonumber \\
&&\ +{\cal B}_{N-1}^{\prime }\otimes \frac 12[({\bf a}_N\cdot {\bf \hat s}%
_N)-({\bf a}_N^{\prime }\cdot {\bf \hat s}_N)],  \label{belln}
\end{eqnarray}
where ${\bf \hat s}_m$ ($m=1$, $2$, $\ldots $, $N$) are the parity spin
operators as defined in Eq. (\ref{sv}), ${\bf a}_m$ and ${\bf a}_m^{\prime }$
are all unit vectors. ${\cal B}_N^{\prime }$ defined in Eq. (\ref{belln})
has the same expression as ${\cal B}_N$, but with ${\bf a}_m\leftrightarrow 
{\bf a}_m^{\prime }$. Similarly to the discrete variable case \cite{Bell-N},
local realistic theories must satisfy the following $N$-party Bell-CHSH
inequality 
\begin{equation}
\left| \left\langle {\cal B}_N\right\rangle \right| \leq 2,  \label{lrn}
\end{equation}
where $\left\langle {\cal B}_N\right\rangle $ is the expectation value of $%
{\cal B}_N$ with respect to any $N$-mode state of the light field.

Meanwhile, an upper bound of $\left| \left\langle {\cal B}_N\right\rangle
\right| $ can be determined quantum mechanically. By analogy to the $N$%
-party Bell-CHSH inequality for discrete variables \cite{Bell-N}, we can
prove, using the properties of ${\bf \hat s}_m$ [see Eq. (\ref{comm})], that 
\begin{eqnarray}
{\cal B}_N^2 &=&{\cal B}_{N-1}^2\otimes \frac 12(1+{\bf a}_N\cdot {\bf a}%
_N^{\prime })  \nonumber \\
&&\ \ \ +{\cal B}_{N-1}^{\prime 2}\otimes \frac 12(1-{\bf a}_N\cdot {\bf a}%
_N^{\prime })  \nonumber \\
&&\ \ \ +[{\cal B}_{N-1},{\cal B}_{N-1}^{\prime }]\otimes \frac i2({\bf a}%
_N\times {\bf a}_N^{\prime })\cdot {\bf \hat s}_N  \nonumber \\
\ &\leq &(1+\left| {\bf a}_N\times {\bf a}_N^{\prime }\right| )\left| {\cal B%
}_{N-1}^2\right| \leq 2^{N+1}.  \label{bound}
\end{eqnarray}
In the last step of Eq. (\ref{bound}), we have used the fact that ${\cal B}%
_2^2\leq 2^{2+1}$ \cite{Chen}, from which the upper bound $2^{N+1}$ of $%
{\cal B}_N^2$ can be recursively obtained. Therefore quantum mechanics gives
an upper bound $2^{(N+1)/2}$ of $\left| \left\langle {\cal B}_N\right\rangle
\right| $; $N$-mode entangled states of the light field can violate the $N$%
-mode Bell-CHSH inequality (\ref{lrn}) by a maximal factor of $2^{(N-1)/2}$,
similarly to the discrete variable case \cite{Bell-N}. For the two-mode
Bell-CHSH inequality, the upper bound $2^{3/2}$ of $\left| \left\langle 
{\cal B}_N\right\rangle \right| $ is known as the Cirel'son bound \cite{MAX}%
. Interestingly, the $N$-party Bell-CHSH inequality (\ref{lrn}), though
expressed in terms of discrete variable operators (i.e., the parity spin
operators), can be exploited to uncover the GHZ nonlocality of continuous
variable states, e.g., the multi-mode entangled states considered in Ref. 
\cite{GHZ-cv}.

Can an $N$-mode entangled state of the light field maximally violate the $N$%
-mode Bell-CHSH inequality (\ref{lrn})? To answer this question, we follow
Mermin's reasoning \cite{Bell-N} and take, as an example, the following $N$%
-mode parity-entangled GHZ state (instead of the multi-mode entangled states
as in Ref. \cite{GHZ-cv}) 
\begin{equation}
\left| {\rm GHZ}\right\rangle _N=\frac 1{\sqrt{2}}\left( \left|
+\right\rangle _1\left| +\right\rangle _2\cdots \left| +\right\rangle
_N-\left| -\right\rangle _1\left| -\right\rangle _2\cdots \left|
-\right\rangle _N\right) .  \label{ghzn}
\end{equation}
From Eq. (\ref{prop}) we have 
\begin{eqnarray}
(s_{mx}\pm is_{my})\left| \mp \right\rangle _m &=&2\left| \pm \right\rangle
_m,  \nonumber \\
(s_{mx}\pm is_{my})\left| \pm \right\rangle _m &=&0.  \label{propxy}
\end{eqnarray}
Then it can be easily verified that $\left| {\rm GHZ}\right\rangle _N$ is an
eigenstate of the following operator 
\begin{eqnarray}
A &=&\frac 12\left[
\prod_{m=1}^N(s_{mx}+is_{my})+\prod_{m=1}^N(s_{mx}-is_{my})\right] , 
\nonumber  \label{a} \\
&&\ \left. A\left| {\rm GHZ}\right\rangle _N=-2^{N-1}\left| {\rm GHZ}%
\right\rangle _N\right.  \label{a}
\end{eqnarray}
with eigenvalue $-2^{N-1}$. Hence, quantum mechanics predicts that 
\begin{equation}
\left| _N\left\langle {\rm GHZ}\right| A\left| {\rm GHZ}\right\rangle
_N\right| =2^{N-1}.  \label{qma}
\end{equation}

By contrast, local realistic theories predict a much lower bound on the
measured results of $A$ for large $N$. Under the notion of local realism,
there are predetermined values $p_{1x}$, $p_{1y}$, $p_{2x}$,$\ldots $ ($%
p_{jx},p_{jy}=\pm 1)$ corresponding to operators $s_{1x}$, $s_{1y}$, $s_{2x}$%
,$\ldots $. Each of these predetermined values of these operators does not
depend on other measurements. Consequently, local realistic prediction on
the upper bound of $A$ reads 
\begin{equation}
\left| {\rm Re}\prod_{m=1}^N(p_{mx}+ip_{my})\right| \leq \left|
\prod_{m=1}^N(\pm 1\pm i)\right| =2^{N/2}.  \label{lra}
\end{equation}
Compared with the quantum-mechanical prediction (\ref{qma}), local realism
predicts a bound of $A$ lower by a large factor $2^{N/2-1}$, growing
exponentially with $N$.

As comparison, within the BW formalism the contradictions between quantum
mechanics and local realism also grow with increasing number of parties for
multi-mode entangled states generated from squeezed states and beam
splitters \cite{GHZ-cv}. But the growth seems to decrease for larger number
of parties, instead of the exponential growth. In this aspect, the merit of
the present formalism is manifest.

To summarize, we have discussed the GHZ nonlocality for continuous
variables, as a natural development of our previous work \cite{Chen}. The
introduction of the parity spin operators satisfying the same algebra as the
Pauli operator is crucial to our discussion. Then the $N$-party Bell
operator of continuous variables was constructed in terms of the parity spin
operators of the $N$-mode light field, resulting in a generalization of the
multiparty Bell-CHSH inequality to the continuous variable cases. Following
Mermin's argument, it has been shown that for $N$-mode parity-entangled GHZ
states (in an infinite-dimensional Hilbert space) of the light field, the
contradictions between quantum mechanics and local realism grow
exponentially with $N$, similarly to the usual $N$-spin cases. Within our
formulation, there is a striking similarity between the multiparty Bell
theorem for both discrete and continuous variables. Our result shows that
only partial (i.e., parity spin) information of the whole Hilbert space of
continuous variables is sufficient for the mere purpose of uncovering
quantum nonlocality. On an experimental aspect, however, the problem of
realizing the parity-entangled GHZ states remains open.

{\it Note added}.---After the completion of the work, we became aware of a
related paper by Massar and Pironio \cite{Massar}. These authors considered
the GHZ paradox for continuous variables in terms of modular and binary
variables.

We thank Rob Clifton for bringing Ref. \cite{Clifton} into our attention.
This work was supported by the National Natural Science Foundation of China
under Grants No. 10104014, No. 19975043 and No. 10028406 and by the Chinese
Academy of Sciences.

\end{document}